\documentclass{article}
\usepackage{amsmath,graphicx}
\usepackage{multirow}
\usepackage{float}
\usepackage{lipsum}
\usepackage{graphicx}
\usepackage{amsmath}
\usepackage{amssymb}
\usepackage[preprint]{spconf}

\ninept
\copyrightnotice{\copyright\ IEEE 2018}
\toappear{To appear in {\it Proc.\ ICASSP 2018, April 15-20, 2018, Calgary, Canada}}

\title{Phonetic and Graphemic Systems for Multi-Genre Broadcast Transcription}

\name{Y. Wang, X. Chen, M. J. F. Gales, A. Ragni and J. H. M. Wong\thanks{This research was partly funded under the ALTA Institute, University of Cambridge. Thanks to Cambridge English, University of Cambridge, for supporting this research.}}
\address{Cambridge University Engineering Dept, Trumpington St., Cambridge CB2 1PZ, U.K.\\
Email: \{yw396, xc257, mjfg, ar527, jhmw2\}@eng.cam.ac.uk}
 
\begin{document}

%
%
\maketitle

\begin{abstract}
State-of-the-art English automatic speech recognition systems typically use phonetic rather than graphemic lexicons. Graphemic systems are known to perform less well for English as the mapping from the written form to the spoken form is complicated. However, in recent years the representational power of deep-learning based acoustic models has improved, raising interest in graphemic acoustic models for English, due to the simplicity of generating the lexicon. In this paper,  phonetic and graphemic models are compared for an English Multi-Genre Broadcast transcription task. A range of acoustic models based on lattice-free MMI training are constructed using phonetic and graphemic lexicons. For this task, it is found that having a long-span temporal history reduces the difference in performance between the two forms of models. In addition, system combination is examined, using parameter smoothing and hypothesis combination. As the combination approaches become more complicated the difference between the phonetic and graphemic systems further decreases. Finally, for all configurations examined the combination of phonetic and graphemic systems yields consistent gains.  
\end{abstract}
\section{Introduction}
\label{sec:intro}
Hidden Markov model (HMM) based automatic speech recognition (ASR) systems are typically built using sub-words units, such as phones or graphemes. System performance  depends on an appropriate definition of  sub-word units and the accuracy, and consistency, of decomposing words into these sub-word units. Phonetic lexicons provide a mapping between the orthographic representation of a word, a sequence of letters (graphemes), into a sequence of phones. However, generation of these lexicons requires linguistic knowledge of the target language, which is time-consuming and expensive. On the other hand, graphemic lexicons are attractive as the graphemes are directly used. Moreover, graphemic lexicons can be easily expanded to include out-of-vocabulary (OOV) words, unlike phonetic lexicons.
For languages with a close grapheme-to-phone mapping, graphemic HMM-based systems have been shown to perform similarly to phonetic systems \cite{Kanthak2002,Killer2003,Gales2015}. However, for languages with irregular grapheme-to-phone mappings, such as English, graphemic HMM-based systems normally perform significantly worse than their phonetic counterparts \cite{Killer2003}. This is not surprising as the system
relies on the acoustic model to implicitly capture the irregularities of the graphemic to acoustic realisation. When more powerful deep learning based acoustic models are used, such as connectionist temporal classification (CTC) \cite{Graves2006} which model long-span temporal information, the gap between graphemic systems and phonetic systems is small on a read English task \cite{Eyben2009}.

This paper aims to find out whether recent deep-learning based acoustic models, which also model long-span temporal information, allow HMM-based graphemic systems to perform at the same level of accuracy as phonetic systems for English. 
A range of models are available including long short-term memory (LSTM) networks \cite{Sak2014}, convolutional neural networks \cite{Abdel2014}, time-delay neural networks (TDNN) \cite{Peddinti2015} and bidirectional LSTM networks \cite{Graves2014}. Additionally various layer-wise combination schemes allow the advantages of several models to be leveraged \cite{Sainath2015,Peddinti2017a}. These models also offer flexibility in terms of the span of the temporal information that they can capture. For instance, the interleaved TDNN-LSTM model \cite{Peddinti2017a} extends the temporal span of the LSTM model with a wide window into the future. These models can also be efficiently trained directly from random initialisation by using approaches such as lattice-free maximum mutual information (LF-MMI) estimation. This often results in improved performance over state-level minimum Bayes' risk (sMBR) trained models \cite{Povey2007}. These complex models are likely to have, possibly significant, variations in ASR performance depending on the choice of training hyper-parameters. This variation in system performance  can be taken advantage of using system combination \cite{Wong2017}. This paper will examine two forms of system combination with different complexities and costs. The first is a random ensemble method \cite{Wong2016}, which utilises multiple training runs with different random seeds to produce slightly different yet complimentary systems. The second is model smoothing \cite{Povey2011}, which interpolates a number of intermediate model parameters using weights estimated on a subset of the training data. Finally, graphemic systems, if competitive with the phonetic system, should be complimentary to phonetic systems. 

The rest of this paper is organized as follows. Sections \ref{sec:am_and_lm} and \ref{sec:model_combination} describe graphemic acoustic models and model combination approaches respectively. Section \ref{sec:experiments} details the experiments conducted on an English multi-genre broadcast transcription task with the phonetic and graphemic models as well as using different combination approaches. Finally, conclusions are given in Section \ref{sec:conclusion}.

\section{Graphemic English systems}
\label{sec:am_and_lm}

\subsection{Graphemic lexicon}
\label{ssec:graphemic_lexicon}
At the core of any graphemic system is the graphemic lexicon. For English, it is straightforward to form this from the 26 alphabet letters /\texttt{a-z}/. In addition to these base graphemes, it may also be useful to mark additional attributes such as apostrophes (DA) and abbreviations (DB).
Excerpts from phonetic and graphemic lexicons are:\\
\centerline{
\begin{tabular}{ll}
\multicolumn{2}{c}{{\bf Phonetic Lexicon}}\\
B.B.C.'s & {\tt /b/ /iy/ /b/ /iy/ /s/ /iy/ /z/}\\
information & {\tt /ih /n/ /f/ /ax/ /m/ /ey/ /sh/ /en/}\\
moon & {\tt /m/ /uw/ /n/}\\
the & {\tt /dh/ /ax/}\\
\multicolumn{2}{c}{{\bf Graphemic Lexicon}}\\
B.B.C.'s & {\tt b;DB b;DB c;DADB s}\\
information  & {\tt i n f o r m a t i o n}\\
moon & {\tt m o o n}\\
the & {\tt t h e}\\
\end{tabular}}
From the first entry, the use of abbreviation and apostrophe attributes potentially allows the graphemic system to handle the discrepancy between the pronounced and written forms. The other three examples illustrate situations where graphemic systems may struggle to model letter omission ({\tt 'r'} in information), vowel ({\tt'oo'} in moon), consonant ({\tt 'th'} in the) and vowel-consonant ({\tt 'tio'} in information) recombination. Though some of these issues can be handled using context dependent models, e.g. bi-graphemes and tri-graphemes, for others the length of the context necessary for disambiguation will be prohibitively large. For example, the phonetic lexicon used in Section~\ref{sec:experiments} associates three phones, {\tt /dh/}, {\tt /th/} and {\tt /t/}, with the sound corresponding to the grapheme sequence {\tt 'th'}. This problem is further compounded by the fact that the following grapheme {\tt 'e'} depending on its neighbor is represented by 9 different vowel/consonant phones. The examples given in this section suggest that for graphemic systems, context modelling may be even more important than it is for phonetic systems \cite{Odell95theuse}. 


\subsection{Acoustic Model Structure}
\label{ssec:mode_structure}
Rather than solely relying on acoustic modelling units to handle the intricate grapheme-to-phone rules, it is also possible to examine acoustic models capable of modelling long-span temporal information. In a deep neural network (DNN) acoustic model \cite{Hinton2012}, only a small number of preceding and succeeding frames
are typically used to predict the current state, $s_t$, as is shown in \eqref{eq:TDNN_model} 
\begin{flalign}
\text{(T)DNN:}\,P\left(s_{t}|\mathcal{O}_{1:T}\right) & \approx P\left(s_{t}|\mathbf{o}_{t-\tau^{\left(l\right)}}\,,\ldots,\,\mathbf{o}_{t+\tau^{\left(r\right)}}\right)\label{eq:TDNN_model}\\
\text{LSTM:}\,P\left(s_{t}|\mathcal{O}_{1:T}\right) & \approx P\left(s_{t}|\mathbf{o}_{1}\,,\ldots,\,\mathbf{o}_{t+\tau^{\left(r\right)}}\right)\label{eq:LSTM_model}
\end{flalign}
where the left $\tau^{(l)}$ and right $\tau^{(r)}$ context window sizes are typically less than 10. A TDNN \cite{Peddinti2015} has a more complex structure that enables it to cover a significantly larger number of preceding and succeeding frames without significantly increasing the number of model parameters. For example, the model considered in Section~\ref{sec:experiments} uses $\tau^{(l)}=15$ past and $\tau^{(r)}=10$ future frames. The use of recurrent units in a LSTM network, described in equation~\eqref{eq:LSTM_model}, allows  even longer-span temporal information to be modelled. Note that in practice the past information is typically truncated after some fixed, yet large, number of frames (40 in this work). Furthermore, the TDNN-LSTM model \cite{Peddinti2017a} obtained by interleaving TDNN layers \cite{Peddinti2015} with LSTM layers \cite{Sak2014} increases the context window to 50 frames into the past and 20 frames into the future. In addition to being more powerful classifiers, these advanced deep-learning based acoustic models thus can utilise a significantly longer span of temporal information than that used in  previous work with Gaussian mixture models.


\section{Model Combination}
\label{sec:model_combination}
Training the deep neural network models discussed in Section~\ref{sec:am_and_lm} is a complicated process involving highly non-convex optimisation. There may thus be large variations between the behaviours of intermediate models from iteration to iteration, or between final models when originating from different starting points. 
The latter is likely to be larger when models are trained from different random initialisations using the LF-MMI criterion, as there is no cross-entropy initialisation stage with common targets for all systems. Such variation typically results in the models making different predictions. Depending on the level of useful variation, such diverse predictions may help to resolve confusions. This serves as the basis for various system combination approaches \cite{Fiscus1997,Evermann2000,Xu2011,Wang2015}. 



\subsection{Ensemble Methods}
\label{sec:ensemble_analysis}
A combination of an ensemble of diverse and yet individually accurate systems can often result in significant gains \cite{Dietterich2000}. Common methods to introduce diversity include random parameter initialisation for ASR \cite{Wong2017, Wong2016}, bagging \cite{Breiman1996} and random decision trees \cite{Siohan2005}. Using different random initialisations has been shown to be a simple and efficient approach of introducing diversity \cite{Wong2017,Wong2016}. In \cite{Wong2016}, this method was able to provide significant diversity while keeping a similar performance across the systems. Thus, combining the systems in the ensemble could yield strong gains. A less common method of ensemble generation is to take a number of intermediate models during training and interpolate their parameters
\begin{equation}
\boldsymbol{\Phi}=\underset{m=1}{\overset{M}{\sum}}\alpha_{m}\boldsymbol{\Phi}_{m}
\label{eq:smoothing}
\end{equation}
where $M$ is the number of models, $\boldsymbol{\Phi}_{m}$ represents the parameters of the $m$th model and $\alpha_{m}$ represents its combination weight. This is the idea behind {\it model smoothing} \cite{Povey2011}, designed to reduce unwanted variations during the training. The models are normally selected from the later stages of training using a fixed iteration interval between the selected models (6 in this work). The combination weights are associated with the individual layers and optimised on a subset of training examples. The combination weights are constrained to sum to 1. Though generally it is hard to ensure that the combined model would improve over the final trained model, this paper shows that large performance improvements are possible. 

To measure the diversity of the generated systems, it is possible to use cross word error rate (cWER) \cite{Wong2017}
\begin{equation}
\text{cWER}=\frac{1}{M\left(M-1\right)}\underset{m=1}{\overset{M}{\sum}}\underset{n\ne m}{\overset{}{\sum}}\frac{1}{\sum_{r=1}^{R}\left|\mathcal{W}_{r}^{n}\right|}\underset{r=1}{\overset{R}{\sum}}L\left(\mathcal{W}_{r}^{m},\mathcal{W}_{r}^{n}\right),
\end{equation}
where $\mathcal{W}_{r}^{m}$ represents the 1-best hypothesis of the $r$th utterance, using the $m$th model, and $R$ is the total number of utterances. The cWER  measures how different the 1-best hypotheses are between models and was found to be more correlated with the combination gains than the standard deviation of WERs \cite{Wong2017}.

\subsection{Minimum Bayes Risk Combination}
\label{sec:MBR_combination}
It is only possible to use model smoothing in equation~\eqref{eq:smoothing} for combining iterations of the same model training run. A more general system combination approach is hypothesis-level combination. Examples of this form of approach are:
ROVER \cite{Fiscus1997}; confusion network combination (CNC) \cite{Evermann2000}; and minimum Bayes risk (MBR) combination \cite{Xu2011}. 
In this work MBR combination is used, which finds the word sequence that attempts to minimise the expected WER across the systems being combined \cite{Xu2011}:
\begin{equation}
\widehat{\mathcal{W}}=\underset{\mathcal{W}}{\text{argmin}}\left\{ \underset{m=1}{\overset{M}{\sum}}\lambda_{m}\underset{\mathcal{W}^\prime\in\mathcal{H}}{\sum}P\left(\mathcal{W}^\prime|\mathcal{O}_{1:T};\boldsymbol{\Phi}_{m}\right)L\left(\mathcal{W},\mathcal{W}^\prime\right)\right\},
\end{equation}
where $\lambda_{m}$ are the combination weights, $P\left(\mathcal{W}|\mathcal{O}_{1:T};\boldsymbol{\Phi}_{m}\right)$ is the posterior probability of the word sequence, $\mathcal{W}$, given the observation sequence, ${\mathcal O}_{1:T}$, and the acoustic model, $\boldsymbol{\Phi}_{m}$, $\mathcal{H}$ is a set of hypotheses and $L\left(\mathcal{W},\mathcal{W}^\prime\right)$ represents the Levenshtein distance between two word sequences $\mathcal{W}$ and $\mathcal{W}^\prime$. Though more computationally expensive, MBR combination has been shown to perform better than the ROVER combination and CNC.

\section{Experiments}
\label{sec:experiments}
Experiments were conducted using the data from the 2017 English Multi-Genre Broadcast (MGB-3) challenge. The data was supplied by British Broadcasting Corporation (BBC) and consists of audio from BBC television programmes. The data contains a wide range of genres such as comedy, drama and sports shows. A total of 375 hours of audio data with associated subtitles  is available for acoustic model training. Lightly supervised decoding and selection was used to extract 275 hours for training \cite{Bell2015,Woodland2015}. A 6 hours development set, dev17b, was also supplied. 
The acoustic model features were 40-dimensional Mel-filter bank features normalised using utterance level  mean normalisation and show-segment level variance normalisation \cite{Woodland2015}. Around 3600 left bi-phone dependent states were used as targets. The results are based on automatic audio segmentation using a DNN based segmenter \cite{Wang2016a} trained on the MGB-3 data.

To examine the impact of the acoustic model complexity on  phonetic and graphemic system performance, a range of acoustic models of different topology and spans of temporal information were built. These include feed-forward DNN, sub-sampled TDNN, unidirectional LSTM and interleaved TDNN-LSTM models. The DNN models had 7 hidden layers of 600-dimensional sigmoid units and an input context window spanning from 10 frames into the past to 10 frames into the future. The TDNN models had 7 layers of 600-dimensional rectified linear units (ReLU) and wider input context window spanning from 15 frames into the past to 10 frames into the future.\footnote{The splicing indexes per layer can be described as \{-1,0,1\} \{-1,0,1\} \{-1,0,1,2\} \{-3,0,3\} \{-3,0,3\} \{-6,-3,0\} \{0\} using the notation of \cite{Peddinti2015,Peddinti2017a}.}
The LSTM model had 3 LSTMP layers, each with 512-dimensional cells and 128-dimensional recurrent and non-recurrent projections. The effective temporal information window for the LSTM spans from 40 frames into the past to 7 frames into the future. The interleaved TDNN-LSTM models had 9 layers of 600 dimensional ReLU units.\footnote{The architecture can be described as \{-2,-1,0,1,2\} \{-1,0,1\} $\mathcal{L}$ \{-3,0,3\} \{-3,0,3\} $\mathcal{L}$ \{-3,0,3\} \{-3,0,3\} $\mathcal{L}$, where $\mathcal{L}$ represents an LSTMP layer with 512 cells and 128-dimensional recurrent and non-recurrent projections, using notation of \cite{Peddinti2015,Peddinti2017a}.} The TDNN-LSTM model has the widest temporal information window, starting from 50 frames into the past and ending at 20 frames into the future. 
All models were trained using the LF-MMI criterion on a single GPU \cite{Povey2016} using Kaldi toolkit \cite{Povey2011}. For this work, only speaker-independent systems were used.

For the first pass decoding language model, a 3-gram language model with a 64K words lexicon was used. This was trained on the audio subtitles and  650M words of supplied BBC subtitles. In addition, a recurrent neural network language model (RNNLM) \cite{Mikolov2010} was also used to refine the result of the first pass decoding. The CUED-RNNLM Toolkit v1.0 \cite{CUEDRNNLM} was used to train the RNNLM using 1 layer of 1024-dimensional GRU units. Given the  vocabulary size (64K) and quantity of training data (e.g. 650M words), noise contrastive estimation (NCE) was adopted to speed up training and evaluation \cite{chen2016taslp}. At test time, a 4-gram approximation \cite{liu2016taslp} of the RNNLM was used to rescore 4-gram lattices. As the RNNLM was trained with the NCE, the unnormalized output layer probabilities were used in rescoring, which provided a large speed up.  MBR decoding/combination was used to produce the final output. Unless stated otherwise, performance with the 3-gram model is quoted. 

\subsection{Phonetic and Graphemic Models}
\label{ssec:Ph_Gr_Models}
\begin{table}[htbp]
\begin{centering}
\begin{tabular}{|cc||cc||cc|}
\hline 
\multicolumn{2}{|c||}{\multirow{2}{*}{Model}} & \multicolumn{2}{c||}{Single} & \multicolumn{2}{c|}{Ph/Gr Comb}\tabularnewline
\cline{3-6} 
 &  & \%WER & \%Rel & \%WER & \%Rel\tabularnewline
\hline 
\hline 
\multirow{2}{*}{DNN} & Ph & 27.8 & \multirow{1}{*}{---} & \multirow{2}{*}{26.3} & \multirow{2}{*}{-5.4}\tabularnewline
 & Gr & 30.7 & +10.4  &  & \tabularnewline
\hline 
\multirow{2}{*}{TDNN} & Ph & 24.4 & \multirow{1}{*}{---} & \multirow{2}{*}{23.0} & \multirow{2}{*}{-5.7}\tabularnewline
 & Gr & 26.9 & +10.3 &  & \tabularnewline
\hline 
\multirow{2}{*}{LSTM} & Ph & 25.0 & \multirow{1}{*}{---} & \multirow{2}{*}{23.2} & \multirow{2}{*}{-7.2}\tabularnewline
 & Gr & 26.7 & +6.8 &  & \tabularnewline
\hline 
\multirow{2}{*}{TDNN-LSTM} & Ph & 23.4 & \multirow{1}{*}{---} & \multirow{2}{*}{21.7} & \multirow{2}{*}{-7.3}\tabularnewline
 & Gr & 25.0 & +6.8 &  & \tabularnewline
\hline 
\end{tabular}
\par\end{centering}
\caption{\%WER of phonetic and graphemic systems and their MBR combination on dev17b. 
\label{tab:WER-different-types}}
\end{table}
The impact of the acoustic model on the performance difference between phonetic (Ph) and graphemic (Gr) systems is illustrated in Table~\ref{tab:WER-different-types}. The second column shows the relative degradation in performance of the graphemic system. As  the complexity of the model and the span of available temporal information increases, the difference between phonetic and graphemic system WERs drops from 10.4 to 6.8\% relative. The largest drop happens when the LSTM units are used to model longer history information. This implies that graphemic systems are more sensitive to shorter histories than are phonetic systems. The third column  in Table~\ref{tab:WER-different-types} shows that as the graphemic system gets more competitive, the gain from combining it with the phonetic system increases from 5.4 to 7.3\% relative. 

\begin{table}[H]
\begin{center}
\begin{tabular}{|cl||c|c|}
\hline 
\multirow{2}{*}{Model} & \multirow{2}{*}{Context} & \multirow{2}{*}{\%WER} & \multirow{2}{*}{RTF}\tabularnewline
 &  &  & \tabularnewline
\hline 
\hline 
\multirow{2}{*}{Ph} & Bi-phone & 23.4 & 0.9\tabularnewline
 & Mono-phone & 23.9 & 0.7\tabularnewline
\hline 
\multirow{2}{*}{Gr} & Bi-grapheme & 25.0 & 0.8\tabularnewline
 & Mono-grapheme & 26.2 & 0.6\tabularnewline
\hline 
\end{tabular}
\par\end{center}
\caption{\%WER of context dependent and independent phonetic and graphemic TDNN-LSTM models on dev17b. \label{tab:WER-monographeme}}
\end{table}
Graphemic systems are also expected to be sensitive to the choice of acoustic modelling context. Wider contexts should be more suitable for graphemic systems as they can better account for the mismatch between the orthographic and spoken form. However, shorter contexts are appealing due to their simplicity and speed of training as well as decoding. Table~\ref{tab:WER-monographeme} shows that phonetic systems are significantly more robust when bi-phone units are replaced with mono-phone units. Though mono-grapheme units yield twice as large a degradation as mono-phone units, the simplicity of graphemic lexicons offers an interesting compromise. Both context-independent systems are approximately 25\% faster than their context-dependent counterparts as shown by the real time factor (RTF) in Table~\ref{tab:WER-monographeme}. 


\subsection{Model combination} 
\label{ssec:MC_experiments}
\begin{table}[htbp]
\begin{centering}
\begin{tabular}{|c||c||cc|}
\hline 
Training  & \multirow{2}*{\%WER} & \multicolumn{2}{c|}{Comb}\tabularnewline
Criterion & & \%WER & \%Rel 
\\
\hline 
\hline 
 sMBR & 23.7 & \multirow{2}{*}{\centering{}21.3} &
 \multirow{2}*{-9.0}
 \tabularnewline
 LF-MMI & 23.4 & & \tabularnewline
\hline 
\end{tabular}
\par\end{centering}
\caption{\%WER of sMBR and LF-MMI trained phonetic TDNN-LSTM models and their MBR combination on dev17b.\label{tab:WER-sMBR-LFMMI}}
\end{table}
Rather than combining phonetic and graphemic systems, it is also possible to combine systems from any diverse ensemble as discussed in Section~\ref{sec:model_combination}. One simple way to produce an additional system is to utilise an alternative training criterion such as sMBR training. Table~\ref{tab:WER-sMBR-LFMMI} shows that sMBR training yields a competitive model and combination gains between systems with these criteria is larger than that between phonetic and graphemic systems in Table~\ref{tab:WER-different-types}. This can partly be attributed to the larger performance differences between the phonetic and graphemic systems being combined.

Additional systems can also be generated using simpler approaches. For example, the use of model smoothing does not require another model to be trained. In this work, 20 models with an iteration interval of 6 were taken from the final epoch of LF-MMI training and their combination weights were estimated on a subset of training data as discussed in Section~\ref{sec:model_combination}. Table~\ref{tab:WER-smoothing} shows that model smoothing is an effective way to improve system performance for both graphemic and phonetic systems. Additionally by performing model smoothing the difference between the phonetic and graphemic systems is reduced (+6.0\%). Though the gains from combining phonetic and graphemic systems decrease after model smoothing, dropping from 7.3 to 5.6\% relative,
there is still a large gain in performance in the combined systems after model smoothing, yielding a better performance than modifying the training criterion.

\begin{table}[H]
\begin{centering}
\begin{tabular}{|c||cc||cc|}
\hline 
\multirow{2}{*}{Model} & \multicolumn{2}{c||}{\%WER} &
\multicolumn{2}{c|}{Ph/Gr Comb} \\
& \multirow{1}{*}{Ph} & \multirow{1}{*}{Gr} & 
\%WER & \%Rel \tabularnewline
\hline 
\hline 
\multicolumn{1}{|c||}{Single} & 23.4 & 25.0 & 21.7 & -7.3 \tabularnewline
Smooth & 21.5 & 22.8 & 20.3 & -5.6\tabularnewline
\hline 
\end{tabular}
\par\end{centering}
\caption{\%WER of phonetic and graphemic TDNN-LSTM models
with and without model smoothing on dev17b.\label{tab:WER-smoothing}}
\end{table}
Alternatively, random ensembles \cite{Wong2016} can be built by changing the random seed used to initialise models for LF-MMI training. This is more expensive than model smoothing, but allows additional diversity to be introduced. LF-MMI training may benefit more from random ensemble generation, as it avoids the cross-entropy initialisation stage of approaches such as sMBR training, where the same targets are normally used for all system, possibly reducing the diversity of the final systems after sequence training. In this work, an ensemble of 3 TDNN-LSTM models was created by building 2 additional models using different seeds for random parameter initialisation. Table~\ref{tab:WER-ensemble} shows that although the WER standard deviation across systems is small, the cWER is large suggesting that these systems may be complementary. To put the cWER number in context, an ensemble of sMBR trained models on the AMI IHM task with a mean WER of 25\% had a cWER of 12\%. The last  block in Table~\ref{tab:WER-ensemble} shows that  ensemble combination of multiple single models yields the large gains of the approaches examined in this work. Given the large gains from model smoothing, it is interesting to examine ensembles of smoothed models. These are also shown in Table~\ref{tab:WER-ensemble}. As expected the cWERs for the ensembles are reduced, as model smoothing reduces the diversity from the precise stopping points. However, there are still large gains of over 7\% relative from the ensemble combination. Additionally, the difference between phonetic and graphemic, smoothed or unsmoothed, systems when combining random ensembles has been reduced to just 5\% relative. 

\begin{table}[htbp]
\begin{centering}
\begin{tabular}{|cc||cc||c||cc|}
\hline 
\multicolumn{2}{|c||}{\multirow{2}*{Model}} & \multicolumn{2}{c||}{\%WER} & \multirow{2}*{\%cWER} & \multicolumn{2}{c|}{Ensemble Comb} \\
& & 
\multirow{1}{*}{$\mu$} & \multirow{1}{*}{$\sigma$} &  & 
\%WER & \%Rel 
\tabularnewline
\hline 
\hline 
\multirow{2}{*}{Ph} & Single & 23.5 & 0.06 & 17.9 & 20.9 & -11.1 \tabularnewline
 & Smooth & 21.6 & 0.10 & 13.5 & 20.0 & -7.4 \tabularnewline
\hline 
\multirow{2}{*}{Gr} & Single & 25.0 & 0.10 & 20.4 & 22.1 & -11.6 \tabularnewline
 & Smooth & 22.8 & 0.06 & 14.9 & 21.0 & -7.9 \tabularnewline
\hline 
\end{tabular}
\par\end{centering}
\caption{\%WER of phonetic and graphemic random ensembles of TDNN-LSTM models with and without model smoothing on dev17b.\label{tab:WER-ensemble}}
\end{table}
Given the small difference between the phonetic and graphemic ensembles, additional gains from combining the systems might be expected. However, the extensive use of combination techniques means that diversity between these ensembles has already been significantly reduced. Table~\ref{tab:final_combination} shows that combining phonetic and graphemic ensembles yields only 0.5\% absolute or 2.5\% relative reduction in WER~\footnote{It is worth noting that the performance of combining an ensemble of two phonetic systems was 20.2\%. Thus simply enlarging the size of the  phonetic ensemble is not expected to match this graphemic/phonetic ensemble performance.}. At this point, it is interesting to see if improved language modelling approaches can yield further benefits. The last column in Table~\ref{tab:final_combination} shows that 4-gram LM rescoring reduces the WER from 19.5 to 18.8\%. The RNNLM gave an additional improvement yielding a final error rate of 17.9\% on this task. 
\begin{table}[htbp]
\begin{centering}
\begin{tabular}{|cc||c||ccc|}
\hline 
\multicolumn{2}{|c||}{\multirow{2}*{Model}} & \multirow{1}{*}{\centering{}Comb} & \multicolumn{3}{c|}{Ph/Gr Comb} \\
\cline{4-6}
& & tg & \multirow{1}{*}{tg } & fg & +rnn\tabularnewline
\hline 
\hline 
\multirow{1}{*}{Ph} & \multirow{2}{*}{Ensemble} & 20.0 & \multirow{2}{*}{\centering{}19.5} & \multirow{2}{*}{\centering{}18.8} & \multirow{2}{*}{17.9}\tabularnewline
Gr &  & 21.0 &  &  & \tabularnewline
\hline 
\end{tabular}
\par\end{centering}
\caption{\%WER of the final MGB-3 system on dev17b. \label{tab:final_combination}}
\end{table}

\section{Conclusion} 
\label{sec:conclusion}
This paper has investigated whether the recent advances in deep learning based approaches have enabled graphemic English ASR systems to reach the performance level of traditionally used phonetic systems. It was found that a combination of long-span temporal history and future information with context-dependent graphemic units is important to obtain competitive performance for graphemic English ASR systems. The relative difference between phonetic and graphemic systems can be further reduced by employing system combination approaches, model smoothing and random ensemble methods were both found to be effective. The combination of these two methods yielded a graphemic English ASR system for multi-genre broadcast transcription that is only 5\% relatively worse than an equivalent phonetic English ASR system, and is complementary. 


\bibliographystyle{unsrt}
\bibliography{Refs.bib}
\end{document}